# Recombinant transfer in the basic genome of *E. coli*

Purushottam Dixit [a,b,1], Tin Yau Pang [a,c,1], F. William Studier [a] and Sergei Maslov [a,d]

[a] Biological, Environmental & Climate Sciences Department, Brookhaven National Laboratory, Upton, New York 11973

[b] Current affiliation: Columbia University Department of Systems Biology, 1130 St. Nicholas Avenue, 8th Floor, New York, NY 10032

[c] Current affiliation: Institute for Bioinformatics, Heinrich-Heine-Universität Düsseldorf, Universitätsstraße 1, 40221 Düsseldorf, Germany

[d] Current affiliation: Bioengineering Department, University of Illinois at Urbana-Champaign, Urbana, IL 61801

[1] These authors contributed equally to this work

Corresponding author: F. William Studier, Biological, Environmental & Climate Sciences Department, Brookhaven National Laboratory, Upton, New York 11973.
Telephone 631-344-3390; studier@bnl.gov

Corresponding author: Sergei Maslov, Telephone 631-344-3742; ssmaslov@gmail.com

Abstract

An approximation to the ~4 Mbp basic genome shared by 32 strains of *E. coli* representing six evolutionary groups has been derived and analyzed computationally. A multiple-alignment of the 32 complete genome sequences was filtered to remove mobile elements and identify the most reliable ~90% of the aligned length of each of the resulting 496 basic-genome pairs. Patterns of single-bp mutations (SNPs) in aligned pairs distinguish clonally inherited regions from regions where either genome has acquired DNA fragments from diverged genomes by homologous recombination since their last common ancestor. Such recombinant transfer is pervasive across the basic genome, mostly between genomes in the same evolutionary group, and generates many unique mosaic patterns. The six least-diverged genome-pairs have one or two recombinant transfers of length ~40-115 kbp (and few if any other transfers), each containing one or more gene clusters known to confer strong selective advantage in some environments. Moderately diverged genome-pairs (0.4-1% SNPs ) show mosaic patterns of interspersed clonal and recombinant regions of varying lengths throughout the basic genome, whereas more highly diverged pairs within an evolutionary group or pairs between evolutionary groups having >1.3% SNPs have few clonal matches longer than a few kbp. Many recombinant transfers appear to incorporate fragments of the entering DNA produced by restriction systems of the recipient cell. A simple computational model can closely fit the data. Most recombinant transfers seem likely to be due to generalized transduction by co-evolving populations of phages, which could efficiently distribute variability throughout bacterial genomes.

Significance Statement

A significant fraction of the length of *E. coli* genomes comprises mobile elements integrated at various sites in a ~4 Mbp basic genome shared by the species. We find that the entire basic genome is continually exchanged by homologous recombination with genome fragments acquired from other genomes in the population. Evolutionary groups appear to exchange DNA preferentially within the same group but also with other groups to different extents. Entering DNA is often fragmented by restriction systems of the recipient cell, with surviving pieces replacing homologous parts of the recipient chromosome. Co-evolving populations of phages that package genome fragments and deliver them to cells that have appropriate receptors are likely mediators of most DNA transfers, distributing variability throughout the species.

\body

The increasing availability of complete genome sequences of many different bacterial and archaeal species, as well as metagenomic sequencing of mixed populations from natural environments, has stimulated theoretical and computational approaches to understand mechanisms of speciation and how prokaryotic species should be defined (1-8). Much genome analysis and comparison has been at the level of gene content, identifying core genomes (the set of genes found in most or all genomes in a group) and the continually expanding pan-genome. Population genomics of *E. coli* has been particularly well studied because of its long history in laboratory research and because many pathogenic strains have been isolated and completely sequenced (9-14). Proposed models of how related groups or species form and evolve include isolation by ecological niche (7-9, 11, 15), decreased homologous recombination as divergence between isolated populations increases (2-4, 8, 14, 16), and co-evolving phage and bacterial populations (6).

*E. coli* genomes are highly variable, containing an array of phage-related mobile elements integrated at many different sites (17), random insertions of multiple transposable elements (18), and idiosyncratic genome rearrangements that include inversions, translocations, duplications, and deletions. Although *E. coli* grows by binary cell division, genetic exchange by homologous recombination has come to be recognized as a significant factor in adaptation and genome evolution (9, 10, 19). Of particular interest has been the relative contribution to genome variability of random mutations (single base-pair differences referred to as SNPs) and replacement of genome regions by homologous recombination with fragments imported from other genomes (here referred to as recombinant transfers or transferred regions). Estimates of the rate, extent and average lengths of recombinant transfers in the core genome vary widely, as do methods for detecting transferred regions and assessing their impact on phylogenetic relationships (12-14, 20, 21).

In a previous comparison of complete genome sequences of the K-12 reference strain MG1655 and the reconstructed genome of the B strain of Delbrück and Luria referred to here as B-DL, we observed that SNPs are not randomly distributed among 3620 perfectly matched pairs of coding sequences but rather have two distinct regimes: sharply decreasing numbers of genes having 0, 1, 2 or 3 SNPs, and an abrupt transition to a much broader exponential distribution in which decreasing numbers of genes contain increasing numbers of SNPs from 4 to 102 SNPs per gene (22). Genes in the two regimes of the distribution are interspersed in clusters of variable lengths throughout what we referred to as the basic genome, namely, the ~4 Mbp shared by the two genomes after eliminating mobile elements. We speculated that genes having 0-3 SNPs may primarily have been inherited clonally from the last common ancestor whereas genes comprising the exponential tail may primarily have been acquired by horizontal transfer from diverged members of the population.

The current study was undertaken to extend these observations to a diverse set of 32 completely sequenced *E. coli* genomes and to analyze how SNP distributions in the basic genome change as a function of evolutionary divergence between the 496 pairs of strains in this set. We have taken a simpler approach than those of Touchon *et al.* (13), Didelot *et al.* (14) and McNally *et al.* (21), who previously analyzed multiple-alignments of complete genomes of *E. coli* strains. The appreciably larger basic genome derived here is not restricted to protein-coding sequences and retains positional information.

## Results and Discussion

**Deriving basic genomes.** We selected for analysis the completely sequenced chromosomal genomes of 32 independently isolated *E. coli* strains from six previously defined evolutionary groups: A, B1, B2, D1, D2 and E (Fig. 1). Whole-genome multiple-alignment of all 32 genomes was produced by the Mauve program (23). Computational filters and procedures described in Materials and Methods generated a 3,955,192-bp alignment that has eliminated essentially all mobile elements and approximates the basic genomes of these 32 strains. The organization of 21 of these 32 basic genomes is that of the comprehensively annotated K-12 laboratory strain MG1655. The remaining 11 genomes contained idiosyncratic inversions and translocations, most of which were reconfigured manually to align with the consensus organization in the multiple-alignment (SI Materials and Methods).

**Reliable genome-wide SNP densities.** The filtered multiple-alignment contains 105 ordered alignment blocks that were arbitrarily divided into tandem strings of 1-kbp segments, starting at the left end of each block. This process generated 3903 segments of 1 kbp in 88 strings from 1 to 370 segments long, separated by 105 shorter right-end segments covering ~1.3% of the multiple-alignment. The segments are numbered 1-4008 from left to right. Segment ends are indexed to both the basic and complete genome sequences of each strain to allow easy comparisons and link to annotations. SNPs between each of the 496 pairs of basic genomes and cumulatively across all 32 strains are extracted directly from the filtered multiple-alignment.

SNPs are accurately identified in regions where all 32 basic genomes are unambiguously aligned, but highly variable regions, particularly where alignment lengths differ, can be problematic. To minimize erroneous SNPs due to such multiple-alignment difficulties, we focused most of our analyses on a set of 3769 segments of 1 kbp that have an approximately normal distribution over cumulative SNP densities of 0.3-18.0% (averaging 7.5%) and cover 95.3% of the filtered 32-genome multiple-alignment (Fig. S1). The 134 most diverged 1-kbp segments in the scattered tail of the distribution extend to 60.2% cumulative SNP density and are primarily in known regions of high variability and subject to known selective pressures, including genes for making O-antigens, LPS (lipopolysaccharides), flagella, fimbrial-adhesins, DNA modification and restriction enzymes, and surface receptors for phages and colicins. The most variable parts of such exchangeable regions did not pass the computational filters and are not present in the aligned basic genomes analyzed here.

Even in regions where cumulative SNP densities are in the normal range, alignment problems involving group-specific or individual deletions, misplaced remnants of IS elements, variable numbers of repeats or other idiosyncrasies can generate false SNPs in some genome-pairs. To minimize such problems, most of our analyses were limited to perfectly aligned 1-kbp segments having no indels. Perfectly aligned segments in the set of 3769 usually cover ~90% of an aligned basic-genome pair. This additional filter reduces average SNP density by 5-15% (but as much as 30% between closely related genomes with few total SNPs).

**SNP distributions in genome-pairs.** Our measure of evolutionary divergence is SNP density, the average percentage of SNPs between perfectly aligned 1-kbp segments in the set of 3769, referred to here as $\Delta$ for an entire basic-genome pair and $\delta$ for individual segments. Distributions of SNP densities among individual segments are shown in Fig. 2a for five basic-genome pairs over the range of $\Delta = 0.38\text{-}2.54\%$. The alignments are between the K-12 reference genome MG1655 and two other group-A genomes, and between MG1655 and one genome each from groups B1, E and B2. The two group-A pairs show a sharply decreasing number of segments with increasing numbers of SNPs from 0-3 per 1-kbp segment (0-0.3% SNP density), which we refer to as the clonal peak on the assumption that most of the segments in that peak are likely to have been clonally inherited from a common ancestor. Consistent with our previous observations using matched protein-coding sequences between MG1655 and B-DL (22), the more highly diverged segments are distributed in a roughly exponential tail extending from the clonal peak.

A clonal peak is apparent only when both of the paired genomes are in the same evolutionary group, but not all genome-pairs within a group show a clonal peak: the 28 pairs between the 8 genomes of group A, the 28 pairs between the 8 genomes of group B1, and the single pairs in D2 and E all show at least a small peak; however, only 4 of the 45 pairs between the 10 genomes of B2 show a clonal peak, and neither the single pair in D1 nor any of the 392 pairs between genomes from different groups show a pronounced clonal peak (Dataset S1). As the clonal peak decreases, the increasing number of segments in the exponential tail maintain approximately the same slope, and the most highly diverged genome-pairs have a broad maximum around $\delta = 1\text{-}2\%$ (Fig. 2b). Our computational model of genome divergence, summarized in a later section and detailed in SI, fits the observed distributions quite well over a broad range of $\Delta$ (Fig. 2b).

**Recombinant transfers and mosaic genomes.** For simplicity we refer to all perfectly aligned 1-kbp segments having 0-3 SNPs as clonal, because distinguishing whether such segments were inherited vertically from a common ancestor or represent incidental matches to mosaic regions acquired by recombinant transfer is not always unambiguous. We also refer to segments having more than 3 SNPs as transferred, meaning that the SNPs were acquired at least in part by recombinant transfer from a diverged genome, even though we recognize that some isolated segments containing 4-6 SNPs are likely to be clonal. Using these designations, we calculated four quantities for each of the 496 basic-genome pairs: $\Delta$, the average SNP density for all perfectly aligned 1-kbp segments from the set of 3769; $f_c$, the fraction of these segments containing 0-3 SNPs, referred to as the clonal fraction; $\Delta_c$, the average SNP density in the clonal fraction; and $\Delta_t$, the average SNP density in the remaining, putatively transferred segments (Dataset S1). The values of $f_c$, $\Delta_c$ and $\Delta_t$ are plotted as a function of $\Delta$ for all 496 genome-pairs in Fig. 3a-c and summarized for all combinations between evolutionary groups in Table S1.

These data show that recombinant transfers are pervasive throughout the basic genome. The 104 genome-pairs in which both genomes are from any one of the six evolutionary groups have clonal fractions that decrease approximately linearly from >0.90 to ~0.15 as $\Delta$ increases from <0.1% to ~1.3% (Fig. 3a). Over the same range, $\Delta_c$ increases from <0.02% to ~0.2% (Fig. 3b). Clearly, most of the divergence between basic-genome pairs within an evolutionary group is due to accumulating recombinant transfers from diverged genomes since their last common ancestor. Assuming random recombinant transfers, a recombining population will generate over time a steady-state population of mosaic genomes with many uniquely different patterns of interspersed clonal and recombinant regions. Examples of mosaic patterns between genome-pairs of different divergence can be visualized in Dataset S2 by scrolling through the spreadsheet, which has color coding and information at the top and bottom to help locate specific features.

**Clonal fraction ($f_c$) and average SNP density in transferred segments ($\Delta_t$) within and between groups.** As clonal fraction decreases with accumulating recombinant transfers, $\Delta_t$ remains relatively constant, averaging 1.39, 1.18 and 1.28% in groups A, B1 and B2 in relatively narrow distributions with slight overlaps (except for considerable scatter in the least-diverged pairs, which have few recombinant transfers) (Fig. 3c, Table S1 and Dataset S1). The average $\Delta_t$ within a group should reflect the average divergence in the recombining population, and the relatively narrow distributions and slight overlaps suggest that each group may be exchanging genome fragments by recombinant transfers primarily within its own recombining population, but occasionally with genomes in other recombining groups. The D1, D2 and E groups are represented by only one genome-pair each. The D1 pair ($f_c = 0.20$) and D2 pair ($f_c = 0.28$) each have more than 2000 recombinant 1-kbp segments and their $\Delta_t$ of 1.58% and 1.66% may be approximately representative of the average divergences in their recombining populations. However, the group-E pair ($f_c = 0.98$) has fewer than 100 recombinant segments and its $\Delta_t$ may be far from representative.

The only inter-group genome-pairs with significant clonal fractions are the 64 pairings between group A and B1 genomes ($f_c = 0.27$-$0.18$ and $\Delta = 1.03$-$1.19$%), similar to the 17 most diverged genome-pairs within the B2 group (those involving SE15 or O127:H6 have $f_c = 0.22$-$0.15$ and $\Delta = 1.03$-$1.22$%). The average of $\Delta_t$ over the 64 A-B1 genome-pairs is 1.41% ± 0.02 (SD) and that of the 17 most diverged B2-B2 pairs is 1.33% ± 0.06. The appreciable clonal fractions in A-B1 genome-pairs suggest that the two groups diverged relatively recently.

The 328 inter-group genome-pairs in the other 14 of the 15 possible pairings between the six evolutionary groups assort into sets of 4 to 80 genome-pairs per combination (Table S1 and Dataset S1). The average values of $\Delta_t$ of all genome-pairs in any single set range from 1.7% to 2.6% and each combination has a very narrow distribution (standard deviation usually less than 0.02). These narrow distributions support the interpretation that the different mosaic genomes in each recombining population are well equilibrated to the average diversity in the population throughout their lengths. The clonal fraction is negligible in all 14 of these sets of inter-group genome-pairs, decreasing from 0.07 to 0.01 as $\Delta$ increases from ~1.6% to 2.6%. As a consequence, $\Delta_t$ is approximately the same as $\Delta$ throughout this range (Fig. 3c). The few 1-kbp segments that appear clonal in the most diverged inter-group genome-pairs are usually highly conserved across the 32 genomes (Dataset S2, set 4) and thus would appear to be clonal whether or not they had been transferred. The longest of these highly conserved regions contains the cluster of 26 ribosomal protein genes *rplQ* to *rpsJ* in 13 tandem 1-kbp segments.

**Six slightly diverged genome-pairs reveal an important mode of recombinant transfer.** The six least-diverged pairs of basic genomes, all of which have a clonal fraction >0.95, have a striking pattern of recombinant transfers. Instead of mosaic transferred regions of various lengths dispersed throughout the basic genome in moderately diverged genome-pairs, each genome-pair has only one or two long transferred regions, each region extending across 42-107 kbp of basic-genome sequence and together containing the vast majority of SNPs attributable to recombinant transfer (Table S2). End points of these transferred regions are even farther apart in the complete genome sequences, ~85-240 kbp, due mostly to mobile elements, which may either have been in the transferred fragment or inserted after acquisition.

These long transferred regions replace the mosaic pattern of the recipient genome region with the mosaic pattern of the homologous region of the donor genome. The number of incidental clonal matches between them and the percentage of each transferred region occupied by clonal matches should decrease with increasing average SNP density in the transferred region. Indeed, the least diverged of the 10 transferred regions in Table S2 (1.37% SNP density) has 15 clonal matches of 1-4 segments covering 23% of the transferred length; the next (1.67% SNP density) has 11 clonal matches constituting 11% of the transferred length; the six transferred regions in the range 2.23-3.57% SNP density have only 1 or 2 clonal matches constituting 2-7% of the transferred length; the transferred region in S88 with 4.05% SNP density has no clonal matches; and the transferred region with $\Delta = 5.27\%$ SNP density has a single clonal segment constituting 1% of its length. The transferred region in the B2 strain S88 came from a group-A genome, as evidenced by tandem strings of clonal segments in this region when the S88 genome is paired with different group-A genomes (Dataset S2, set 4).

Initially, it appeared that short recombinant transfers were present elsewhere in the six genome-pairs containing long transfers. However, examination of the candidate SNP clusters provided other explanations for almost all of them. The most interesting proved to be the result of shuffling of variants in ribosomal RNA operons by internal recombination (gene conversion) among the seven operons characteristic of *E. coli* (Dataset S2). These internally generated SNPs in rRNA operons can be a significant fraction of putative recombinant SNPs in the six least-diverged genome-pairs. Most of the other high-density SNP clusters can be attributed to multiple-alignment difficulties. Internally generated and erroneous SNP clusters become a negligible fraction of transferred SNPs in even moderately diverged genome-pairs and corrections for them were not made in the figures and SI datasets. Three isolated clusters of 28-38% SNPs in 88-197 bp in APEC but not in any of the other 31 genomes are the most likely candidates in the six least-diverged genome-pairs to have resulted from short recombinant transfers (Dataset S2, set 2).

Each of the long transferred regions contains gene clusters known to have exchangeable variants that provide a strong selective advantage in some situations. These variable gene clusters can nonetheless be exchanged by homologous recombination because the sequences flanking them are much less variable. The O-antigen and DNA restriction genes were previously identified as having many variants that are frequently exchanged and efficiently retained in *E. coli* populations because they can confer significant selective advantage (24). An O-antigen gene cluster was transferred into at least one of the genomes in all six pairs, the DNA restriction cluster into two genomes, and a gene cluster for capsule formation in one (Table S2). It seems likely that each of these long recombinant transfers was the initial step in divergence of a new lineage in a population under stress, and that the selective advantage it conferred fixed not only the advantageous gene cluster but also the unique mosaic pattern of the recipient genome as the ancestral sequence of the new lineage.

The last of these selective transfers in each of the nine different genomes in the least-diverged genome-pairs was apparently recent enough that no subsequent relatively neutral recombinant transfers have become fixed in the population since the last common ancestor (with the possible exception of the three short candidates in APEC). Rough estimates of the number of generations since the mosaic patterns in the >95% of basic-genome length in these six genome-pairs were fixed can be made by dividing their corrected numbers of SNPs per bp by twice the estimated mutation rate of $8.9 \times 10^{-11}$ mutations per bp per generation (25) (arbitrarily assigning half of the SNPs to each genome in a pair). This calculation gives estimates of $\sim 8 \times 10^5$ generations for the five B2 genomes and $\sim 3 \times 10^6$ for the two B1 and two E genomes.

**DNA restriction has a prominent role in recombinant transfer.** Many types and specificities of DNA restriction and modification are widely distributed in *E. coli* and about half of completely sequenced *E. coli* genomes contain one or more of the many variants of *hsd* genes, which specify Type I restriction/modification systems, and often other restriction enzymes as well (26, 27). Milkman and colleagues used restriction fragment length polymorphism (RFLP) to show that genome fragments introduced into *E. coli* by transduction or conjugation are fragmented and reduced in length by restriction systems, a process they postulated is responsible for generating the mosaic structure of *E. coli* genomes (28, 29). More recently, we analyzed three completely sequenced genomes to deduce the patterns of recombinant transfer by P1 transduction of genomic DNA from the K-12 strain W3110 across a type-I restriction barrier into two different B genomes (22). P1 is a generalized transducing phage that delivers genome fragments as long as ~115 kbp without accompanying phage DNA (30). One of these recombinant transfers delivered a single fragment of 6.0-10.6 kbp but the second delivered six DNA fragments totaling 44.9-55.1 kbp across 71.3-77.0 kbp of genome. The average length of the six transferred W3110 fragments is ~ 8.3 kbp and that of the five clonal B intervals is ~4.8 kbp but the lengths of individual fragments have very wide possible ranges, 0.3-26.5 kbp for transferred fragments and 0.1-13.9 kbp for clonal intervals. We examined nucleotide sequences of all 32 genomes in the primary locus of genes specifying Type I and other types of restrictions enzymes, referred to as the Immigration Control region (ICR) (31). Eighteen of the 32 genomes appear to have the intact *hsd* genes needed for Type I restriction, with several different types represented (details in SI).

**Lengths of interspersed clonal and recombinant regions.** As random recombinant transfers accumulate in genomes that have a recent common ancestor, average lengths of clonal regions decrease rapidly with the corresponding increases in Δ (Fig. 4a and Dataset S3). The seven least-diverged genome-pairs (adding the group-A pair MG-F18) all have clonal fractions >0.90, uninterrupted strings of clonal segments hundreds of kbp long, and one, two or possibly five recombinant transfer events consistent with entering DNA longer than 40 kbp (Table S2, Datasets S2 and S3). The next levels of divergence are seen in a set of six group-A genome-pairs having clonal fractions between 0.76 and 0.57 ($\Delta$ = 0.38-0.64%), five of which have 20-50% of their aligned lengths in uninterrupted strings of clonal segments longer than 50 kbp (Dataset S3). Including all 4008 segments in the analyses (and allowing clonal regions to extend through the high-SNP segments in rRNA operons, segments containing obviously erroneous SNPs, and rare isolated segments containing 4-6 SNPs per kbp), these moderately diverged genome-pairs have as many as seven clonal regions longer than 100 kbp, the longest covering 967 kbp of aligned length (Dataset S2, set 3).

These long clonal regions are distributed across the paired genomes at intervals consistent with clonal inheritance from a recent common ancestor interrupted by random recombinant transfers of indeterminate numbers and lengths from genomes with different mosaic patterns. However, the P12b genome uniquely has long clonal matches with two different genomes: five clonal regions of 184-715 kbp are evident when P12b is aligned with MG, and a single 756-kbp clonal region when aligned with Crooks (in a region that has a divergent mosaic pattern relative to MG, Dataset S2, set 3). The long clonal region between P12b and Crooks must almost certainly be due to an uninterrupted recombinant exchange of ~20% of the basic genome of P12b with DNA delivered by conjugation from a close relative of Crooks.

At higher divergence, average lengths of clonal matches between genomes within an evolutionary group decrease to 1.5-3 kbp, and the longest clonal regions in moderately to highly diverged pairs within a group are typically less than 1% of the aligned basic-genome length (Fig. 4a and Dataset S3). The lengths and distributions of clonal and transferred regions reflect the uniquely different mosaic patterns generated by random recombinant transfers in the two lineages, and incidental clonal matches in transferred regions should decrease with increasing distance from the last common ancestor of donor and recipient. Fragmentation of the entering DNA in a significant fraction of transfers can also reduce average lengths of both clonal and transferred regions. Average lengths of transferred regions are ~2.6-4.6 kbp in the range of $\Delta$ = 0.38-1.0% before increasing at higher divergence as newly acquired recombinant transfers overlap those acquired previously (Fig. 4b). The average number of interspersed regions reaches a maximum around 500 in group-A genome-pairs and around 600 in B1 and B2 genome-pairs before decreasing somewhat in the B2 pairs having $\Delta$ >1.0 % due to overlaps (Fig. 4c).

**Computational model of divergence.** We developed a simple computational model of divergence in a steady-state population of genomes assumed to comprise 4000 segments of 1 kbp and to be accumulating random point mutations and acquiring random tandem segments of variable lengths by recombinant transfer from other genomes in the population at fixed rates. Analytical derivation of the probabilities of possible outcomes, combined with Markov chain simulations, generated probable SNP distributions across 100 pairs of genomes evolving to any given $\Delta$. The model (details in SI) fits the data quite well (Figs. 2c, 3 and 4) using a mutation rate $\mu = 8.9 \times 10^{-11}$ SNPs per bp per generation (25), an average length of recombinant transfer of 3 kbp (Fig. 4b and Dataset S3), and a ratio of rate of accumulation of SNPs by recombinant transfer relative to random mutations, $\rho/\mu$, of 0.31, slightly higher than the 0.14-0.21 calculated in Dataset S1 for group A, B1 and B2 genome-pairs. The value of r/m, a measure of the ratio of the number of mutations in transferred regions relative to random mutations throughout the genome, is 11.2 when calculated from the parameters used in the model, slightly higher than the 8.9 for group A and 5.4 for groups B1 and B2 in Dataset S1 (overall avg = 6.5, SD 0.2). These values for r/m may be compared with the original estimate of 50 by Guttman and Dykhuizen (20), from limited data, and 0.34 by McNally et al. (21), 1.5 by Touchon et al. (13) and 7 by Didelot et al. (14) from complete genome sequences.

**Co-evolving transducing phages as primary vectors of transfer.** How are the continuing, pervasive, and primarily group-specific transfers of genome fragments of 40-115 kbp (or even larger) accomplished routinely in recombining populations of *E. coli*? Of the three well-studied mechanisms of DNA transfer (conjugation, transformation and transduction), generalized transduction seems to us likely to be the primary mode of routine transfer. Generalized transducing phages such as coliphage P1 (30) are widely distributed and can have packaging capacity as high as 300 kbp of DNA (32-34). Co-evolving populations of phages and bacteria (6) would maintain specificity for delivery of genome fragments primarily to members of the recombining population of bacteria, but host-range variability could provide occasional delivery of genome fragments to cells of other populations. Transducing phage particles potentially deliver random, well-protected genome fragments throughout a co-evolving population much more widely and efficiently than conjugation or conjugative plasmids, which require specialized transfer mechanisms and cell-to-cell contact. That is not to argue that conjugation or transfer of conjugative plasmids does not occur or is not important, and we did detect one obvious example of conjugation in the P12b genome. However, phages are ideally suited to mediate the pervasive and continuing recombinant transfer documented by our analyses, and co-evolution of phage and bacterial populations provides a simple explanation for a large body of previous work on genome variability, recombinant transfer and speciation.

**Basic genome: a platform for annotation.** Further development of the basic-genome platform and computational methodology developed here could simplify and facilitate classification and annotation of the current and anticipated flood of complete, draft and metagenome sequences of *E. coli*. Consensus basic-genome sequences of all 32 strains and of the group A, B1 and B2 strains are given as FASTA files in Datasets S4-7. Alignment to these consensus sequences can help to classify newly sequenced *E. coli* genomes, identify orthologs, and distinguish types and locations of mobile elements. Standardized basic-genome annotations and catalogs of group-specific features, exchangeable gene clusters, and other features could accelerate and improve uniformity and reliability of annotation. The methodology should be applicable to any bacterial or archaeal species or evolutionary group.

## Materials and Methods

**Extracting basic genomes.** We used the Mauve program with default parameters (23) to produce a multiple-alignment of the complete genome sequences of 32 independently isolated strains of *E. coli*. Their names, GenBank accession numbers, complete genome lengths and derived basic-genome lengths are given in Table S3, and a GenBank-style annotated complete genome sequence of B-DL is given as Dataset S8.

A key step in deriving basic genomes from the multiple-alignment was to apply a simple computational filter designed to eliminate mobile elements, idiosyncratic insertions or duplications, and highly diverged regions that do not align well. The filter applied in the present analysis removed every base-pair position in which fewer than 22 of the 32 genomes have an aligned base pair, which retained idiosyncratic and group-specific deletions. This filter reduced the initial 3044 Mauve alignment blocks to 105 ordered blocks and the total aligned sequence within these blocks from 20.5 Mbp to the 3,955,192-bp approximation to the basic genome analyzed here. We determined that all mobile elements (prophages, rhs elements, IS elements) annotated in the extensively analyzed complete genome sequences of MG1655 and B-DL (22) were removed by the filter, but at least a few other genomes retained remnants due to difficulties in multiple-alignment at sites where mobile elements are integrated. Our procedures appear to have captured a reasonable approximation to the basic genome shared by these 32 strains and provide a platform for further analysis and refinement.

**Consensus sequences.** Majority rule at each position where at least 22 genomes are represented generated a consensus basic-genome sequence for all 32 genomes. At least 6 genomes were required at each position in the group A and B1 consensus sequences, and at least 7 genomes in B2.

## Acknowledgements

Work was supported by grants PM-031 and ELS165 from the Office of Biological and Environmental Research of the U.S Department of Energy and internal research funding from Brookhaven National Laboratory.

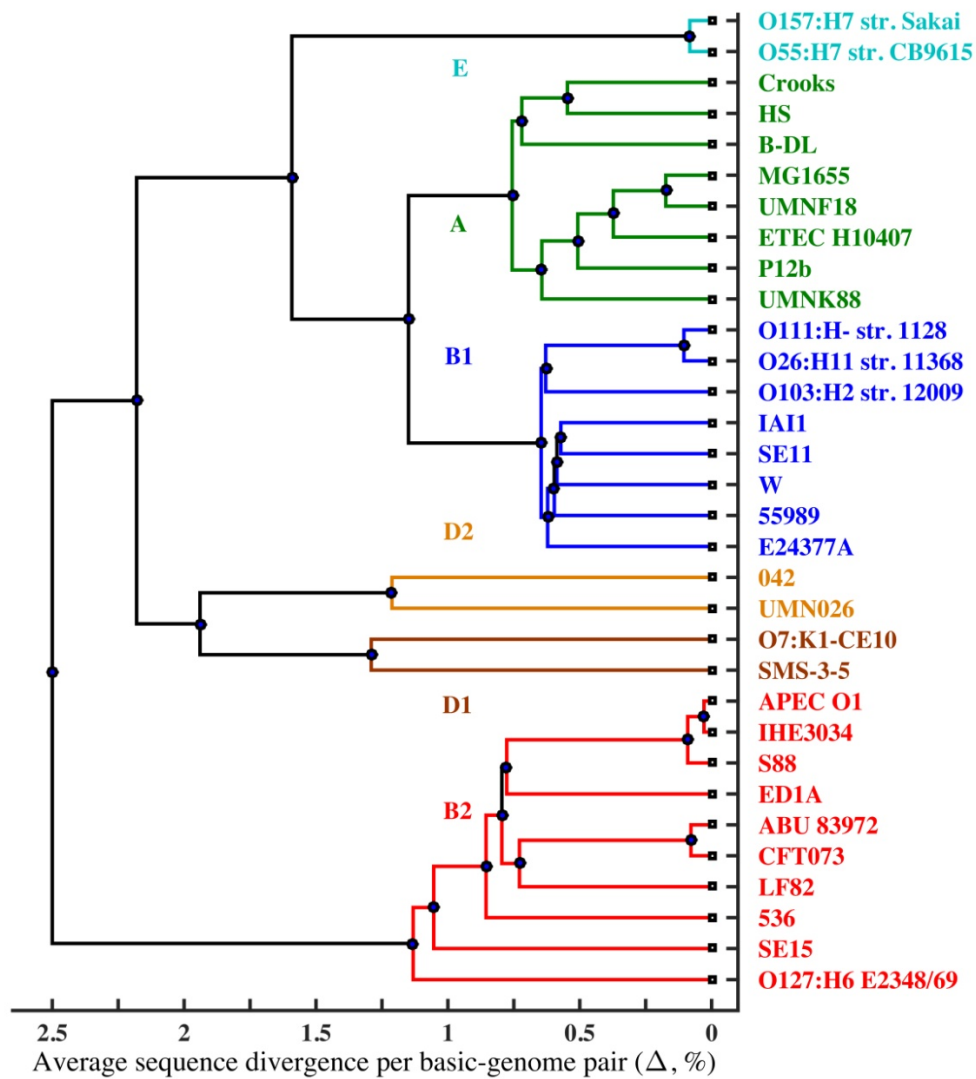

**Figure 1. Phylogenetic tree derived from filtered genome-wide average SNP densities (Δ) between 496 pairs of 32 basic genomes.** Previously recognized phylogenetic groups: E (light blue); A (green); B1 (blue); D2 (yellow); D1 (brown); and B2 (red). The tree was calculated using UPGMA algorithm. Dots in the lines connecting pairs of evolutionary groups are placed approximately at average SNP densities between them. The groups are ordered to fit the relative divergences among them summarized in Table S1. GenBank accession numbers are given in Table S3.

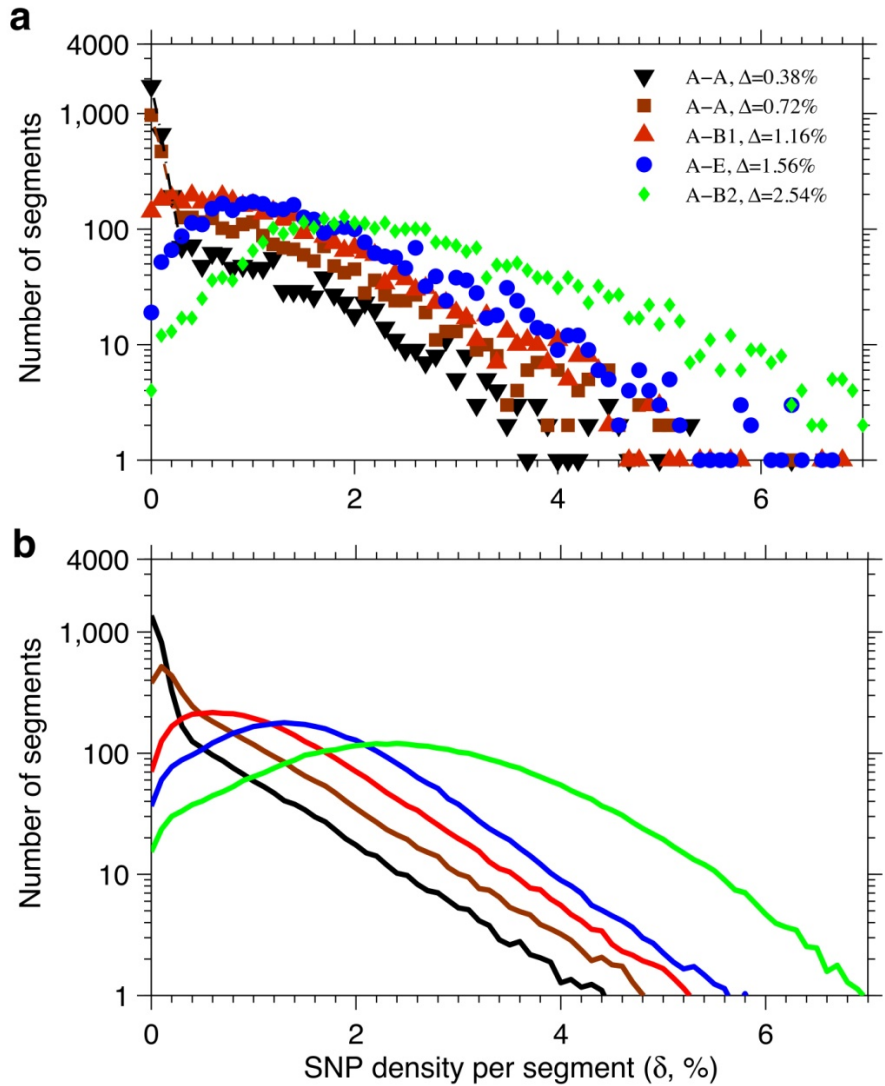

**Figure 2. Distributions of SNP densities between basic genomes. a.** Distribution of perfectly aligned 1-kbp segments from the set of 3769 as a function of average SNP density δ for five basic-genome pairs: group A strain MG1655 aligned with ETEC (A-A, Δ = 0.38%, black triangles); B-DL (A-A, Δ = 0.72%, brown squares); SE11 (A-B1, Δ = 1.16%, red triangles); O157 (A-E, Δ = 1.56%, blue circles); and IHE (A-B2, Δ = 2.54%, green diamonds). **b.** Distributions of SNP density as a function of δ as predicted by the computational model given in SI for pairs of genomes having the same average SNP densities Δ as the five genome-pairs in panel **a**.

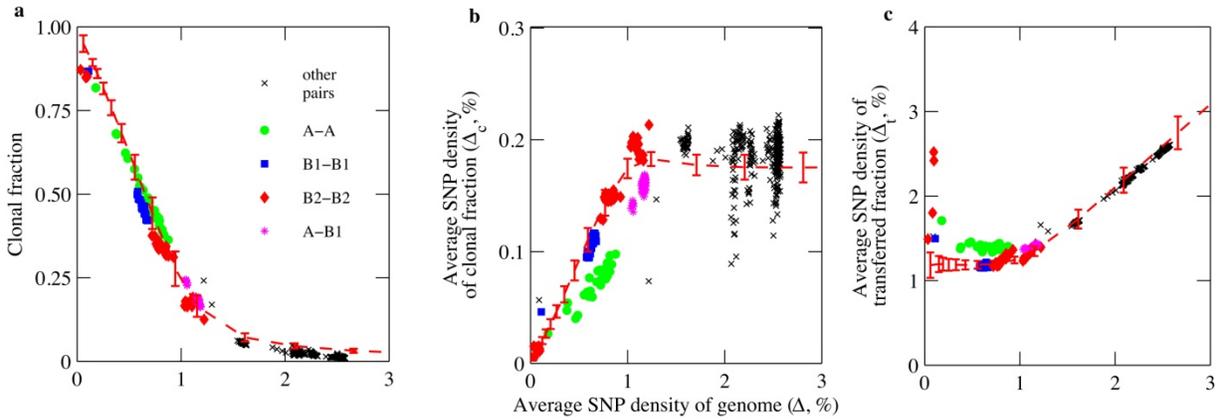

**Figure 3. Values of clonal fraction, $\Delta_c$ and $\Delta_t$ as a function of overall divergence $\Delta$ in 496 basic-genome pairs. a.** Clonal fraction $f_c$; **b.** Average SNP density in the clonal fraction $\Delta_c$; **c.** Average SNP density in the transferred fraction $\Delta_t$. Data points for genome-pairs within evolutionary groups A are given by solid green circles; B1 by blue squares; B2 by red diamonds; between A and B1 by violet asterisks; and other pairs by a black x. Dashed red lines with error bars are values predicted by the computational model given in SI. Error bars correspond to standard deviation of 100 runs of the model.

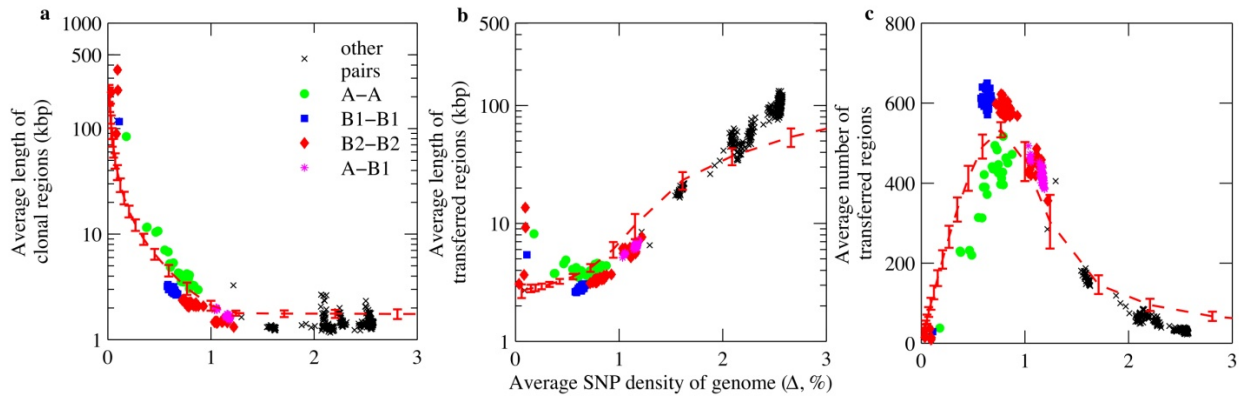

**Figure 4. Average lengths of clonal and transferred regions, and numbers of each as a function of overall divergence $\Delta$ in 496 basic-genome pairs. a.** Average lengths of clonal regions (kbp); **b.** Average lengths of transferred regions (kbp); **c.** Number of transferred regions (equal numbers of interspersed clonal and transferred regions). Data points, dashed red lines and error bars are as in Fig. 3.

# Supporting Information

Table S1. Summary of divergences between 496 basic-genome pairs within and between evolutionary groups, limited to perfectly aligned 1-kbp segments from the set of 3769

| Genome-pairs | | # | Δ min, % | Δ max, % | Clonal fraction max | Clonal fraction min | Δt* avg, % | st dev |
|---|---|---|---|---|---|---|---|---|
| A  | A  | 28 | 0.179 | 0.876 | 0.909 | 0.405 | 1.393 | 0.031 |
| B1 | B1 | 28 | 0.111 | 0.678 | 0.956 | 0.468 | 1.175 | 0.016 |
| B2 | B2 | 45 | 0.034 | 1.221 | 0.987 | 0.147 | 1.279 | 0.071 |
| D1 | D1 | 1  |       | 1.296 |       | 0.200 | 1.583 |       |
| D2 | D2 | 1  |       | 1.217 |       | 0.278 | 1.658 |       |
| E  | E  | 1  | 0.088 |       | 0.979 |       |       |       |
|    |    |    |       |       |       |       |       |       |
| A  | B1 | 64 | 1.034 | 1.189 | 0.271 | 0.184 | 1.413 | 0.016 |
| A  | E  | 16 | 1.542 | 1.593 | 0.071 | 0.060 | 1.676 | 0.013 |
| B1 | E  | 16 | 1.602 | 1.629 | 0.064 | 0.055 | 1.706 | 0.006 |
| D1 | D2 | 4  | 1.879 | 2.008 | 0.049 | 0.031 | 2.018 | 0.044 |
| A  | D2 | 16 | 2.066 | 2.119 | 0.055 | 0.027 | 2.168 | 0.024 |
| B1 | D2 | 16 | 2.072 | 2.130 | 0.031 | 0.021 | 2.151 | 0.022 |
| D2 | E  | 4  | 2.126 | 2.162 | 0.026 | 0.019 | 2.189 | 0.019 |
| B2 | D1 | 20 | 2.111 | 2.195 | 0.040 | 0.024 | 2.228 | 0.017 |
| A  | D1 | 16 | 2.226 | 2.265 | 0.041 | 0.020 | 2.317 | 0.009 |
| B1 | D1 | 16 | 2.251 | 2.295 | 0.028 | 0.018 | 2.323 | 0.010 |
| D1 | E  | 4  | 2.299 | 2.314 | 0.022 | 0.018 | 2.350 | 0.003 |
| B2 | D2 | 20 | 2.392 | 2.512 | 0.017 | 0.012 | 2.499 | 0.030 |
| A  | B2 | 80 | 2.496 | 2.566 | 0.028 | 0.011 | 2.578 | 0.012 |
| B1 | B2 | 80 | 2.516 | 2.575 | 0.016 | 0.011 | 2.581 | 0.012 |
| B2 | E  | 20 | 2.533 | 2.578 | 0.015 | 0.010 | 2.594 | 0.011 |

*excluding genome-pairs with clonal fraction >0.9 in A, B1, B2 and E

Table S2. Long transfers in the least-diverged basic-genome pairs

| Genomes and group | | | Δ, % | $\Delta_c$,% | $f_c$ | %SNPs | Clonal gaps # | Clonal gaps %length | Region | Length kbp ba | Length kbp cg |
|---|---|---|---|---|---|---|---|---|---|---|---|
| IHE | APEC | B2 | 0.034 | 0.015 | 0.99 | 2.23 | 2 | 7.1 | O-antigen | 58 | 102 |
| IHE | S88 | B2 | 0.095 | 0.013 | 0.97 | 2.86 | 2 | 7.1 | O-antigen | 58 | 104 |
| | | | | | | 4.05 | 0 | | rest/fimbri | 107 | 240 |
| APEC | S88 | B2 | 0.097 | 0.011 | 0.96 | 3.34 | 1 | 10.8 | O-antigen | 42 | 85 |
| | | | | | | 4.05 | 0 | | rest/fimbri | 107 | 231 |
| ABU | CFT | B2 | 0.084 | 0.015 | 0.96 | 3.01 | 2 | 3.1 | O-antigen | 65 | 86 |
| | | | | | | 1.37 | 15 | 23.0 | capsule | 104 | 223 |
| O111 | O26 | B1 | 0.111 | 0.046 | 0.96 | 5.27 | 1 | 1.0 | O-antigen | 102 | 202 |
| | | | | | | 1.67 | 11 | 11.3 | rest/fimbri | 105 | 162 |
| O55 | O157 | E | 0.088 | 0.057 | 0.98 | 3.57 | 2 | 2.1 | O-antigen | 94 | 122 |

%SNPs is SNP density in long transferred regions in the total set of 4008 segments, including clonal gaps
rest/fimbri refers to DNA restriction and nearby fimbrial gene clusters
ba is basic genome
cg is complete genome

Table S3. *E. coli* strains in 32-genome Mauve multiple-alignment

|   | Strain | GenBank | Complete genome | Basic genome | ba/cg % | ba/32 % | Reconfigured for alignment |
|---|---|---|---|---|---|---|---|
| A | MG1655 | U00096.2 | 4,639,675 | 3,917,107 | 84.4 | 99.0 |  |
| A | B-DL | B-DL.gb.txt in SI | 4,620,778 | 3,892,824 | 84.2 | 98.4 |  |
| A | HS | CP000802.1 | 4,643,538 | 3,884,968 | 83.7 | 98.2 |  |
| A | P12b | CP002291.1 | 4,935,294 | 3,901,785 | 79.1 | 98.6 | 1.2-Mbp invert btw IS3s |
| A | Crooks (ATCC 8739) | CP000946.1 | 4,746,218 | 3,914,315 | 82.5 | 99.0 | invert and rotate |
| G | UMNF18 | AGTD01000001.1 | 5,239,207 | 3,920,484 | 74.8 | 99.1 |  |
| A | ETEC H10407 | FN649414.1 | 5,153,435 | 3,912,687 | 75.9 | 98.9 |  |
| A | UMNK88 | CP002729.1 | 5,186,416 | 3,927,265 | 75.7 | 99.3 |  |
| B1 | W | CP002185.1 | 4,900,968 | 3,930,428 | 80.2 | 99.4 |  |
| B1 | 55989 | CU928145.2 | 5,154,862 | 3,914,831 | 75.9 | 99.0 | rotate |
| B1 | O26:H11 str. 11368 | NC_013361.1 | 5,697,240 | 3,912,181 | 68.7 | 98.9 | 471-kbp & 5.3-kbp inverts |
| B1 | O111:H- str. 11128 | AP010960.1 | 5,371,077 | 3,910,010 | 72.8 | 98.9 | 92-kbp invert btw IS129s |
| B1 | O103:H2 str. 12009 | NC_013353.1 | 5,449,314 | 3,924,647 | 72.0 | 99.2 | 456-kbp invrt w 5 transloc |
| B1 | SE11 | AP009240.1 | 4,887,515 | 3,932,908 | 80.5 | 99.4 |  |
| B1 | E24377A | CP000800.1 | 4,979,619 | 3,929,451 | 78.9 | 99.3 |  |
| B1 | IAI1 | CU928160.2 | 4,700,560 | 3,927,748 | 83.6 | 99.3 |  |
| E | O55:H7 str. CB9615 | CP001846.1 | 5,386,352 | 3,892,416 | 72.3 | 98.4 |  |
| E | O157:H7 str. Sakai | NC_002695.1 | 5,498,450 | 3,849,037 | 70.0 | 97.3 |  |
| B2 | SE15 | AP009378.1 | 4,717,338 | 3,783,265 | 80.2 | 95.7 | invert between asn tRNAs |
| B2 | IHE3034 | CP001969.1 | 5,108,383 | 3,799,695 | 74.4 | 96.1 | invert between asn tRNAs |
| B2 | S88 | CU928161.2 | 5,032,268 | 3,808,318 | 75.7 | 96.3 |  |
| B2 | APEC O1 | CP000468.1 | 5,082,025 | 3,800,601 | 74.8 | 96.1 | rotate |
| B2 | ED1A | NC_011745.1 | 5,209,548 | 3,739,178 | 71.8 | 94.5 | rotate, 8.8-kbp invert |
| B2 | 536 | CP000247.1 | 4,938,920 | 3,797,593 | 76.9 | 96.0 | invert between asn tRNAs |
| B2 | LF82 | CU651637.1 | 4,773,108 | 3,796,695 | 79.5 | 96.0 | rotate |
| B2 | ABU 83972 | CP001671.1 | 5,131,397 | 3,794,741 | 74.0 | 95.9 | invert between asn tRNAs |
| B2 | CFT073 | AE014075.1 | 5,231,428 | 3,784,222 | 72.3 | 95.7 | invert between asn tRNAs |
| B2 | O127:H6 E2348/69 | FM180568.1 | 4,965,553 | 3,779,984 | 76.1 | 95.6 |  |
| D | O7:K1 str. CE10 | CP003034.1 | 5,313,531 | 3,903,557 | 73.5 | 98.7 |  |
| D | SMS-3-5 | NC_010498.1 | 5,068,389 | 3,911,226 | 77.2 | 98.9 | 1.4-Mbp invrt btw IS110s |
| D | 042 | FN554766.1 | 5,241,977 | 3,879,605 | 74.0 | 98.1 |  |
| D | UMN026 | NC_011751.1 | 5,202,090 | 3,888,373 | 74.7 | 98.3 | rotate |

Length of filtered 32-genome Mauve multiple-alignment:   3,955,192

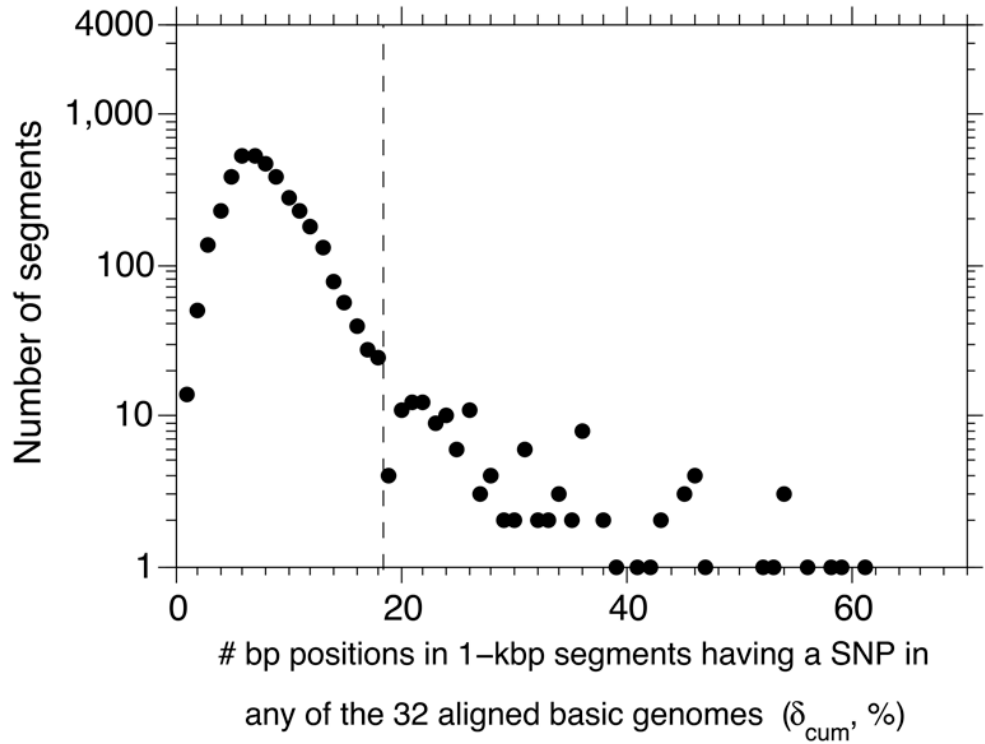

**Fig. S1.** Distribution of the 3903 basic-genome segments of 1 kbp as a function of cumulative SNP density in the filtered 32-strain basic-genome alignment ($\delta_{cum}$ = % of aligned bp positions having a SNP in any strain). An approximately normal distribution of 3769 segments with $\delta_{cum}$ = 0.3-18.0%, average 7.5%, is followed by a scattered tail of 134 segments with $\delta_{cum}$ = 18.1-60.2%, average 29.1%.

# SI Results and Discussion

**DNA restriction.** We previously analyzed three completely sequenced genomes to deduce the patterns of recombinant transfer by P1 transduction of genomic DNA from the K-12 strain W3110 across a type-I restriction barrier into ancestors of two different B genomes, BL21(DE3) and REL606 (22). Transductants had been selected for the repair of a 6-kbp deletion that had apparently been caused in B strains by an IS1 element that remained at the site of the deletion. The only recombinant transfer detected in the BL21(DE3) lineage was a fragment of maximum length 10.6 kbp that replaced the IS1 element and restored the deleted region. However, recombinant transfers in theREL606 lineage replaced six DNA fragments totaling 44.9-55.1 kbp across 71.3-77.0 kbp of genome. Minimum and maximum lengths were determined by the ends of uninterrupted successions of single-bp differences that define which DNA is represented (positions of all single-bp differences among strains are given in Table S1 of reference 22). The observed patterns of recombinant transfers are consistent with the positions of known recognition sites for the B restriction endonuclease in W3110 DNA and the mechanism of action of the B restriction endonuclease (22). The average length of the six transferred W3110 fragments is ~ 8.3 kbp and that of the five clonal B intervals is ~4.8 kbp but the lengths of individual fragments have very wide possible ranges, 0.3-26.5 kbp for transferred fragments and 0.1-13.9 kbp for clonal intervals.

We examined nucleotide sequences of all 32 genomes in the primary locus of genes specifying Type I and other types of restrictions enzymes, referred to as the Immigration Control Region (ICR) (31). Fourteen of these genomes lack all or part of the *hsd* genes needed for Type I restriction, some by exchange of the entire ICR with the unrelated *pac* gene (which specifies a penicillin acylase), some by deletions associated with IS elements, and some by simple deletions. Most of genomes are highly variable relative to each other in at least the specificity gene *hsdS*, if not all three *hsd* genes. However, five pairs of genomes have essentially or completely identical sequences through the entire *hsd* region (and across most or all of the entire ICR), including three of the six least-diverged genome-pairs (IHE-APEC, ABU-CFT and O55-O157). In addition, O111 is one of the genomes that has exchanged the entire ICS with the *pac* gene, and the long transferred region containing the ICR in the O111-O26 genome-pair could either have delivered active *hsd* genes to O26 or replaced them with the *pac* gene in O111.

# SI Materials and Methods

**Extracting basic genomes.** The results given in the figures and most tables are limited to perfectly aligned 1-kbp segments in the set of 3769. Extending the analyses to the entire set of 4008 basic-genome segments or increasing the clonal cutoff from 3 to 4 or 5 SNPs per 1-kbp segment produced similar results with relatively minor differences or trends.

Strain names of the 32 genomes used in this work, GenBank accession numbers, the lengths of both complete genomes and the basic genomes derived from them, and summaries of reconfigurations of the genome sequences made to simplify the multiple-alignment are given in

Table S3. No significant inversions or translocations are seen among 21 of the 32 genomes, including the reference genome MG1655. The first bp in the numbering of six genome sequences was located at a different position relative to gene content from that of MG1655, and those sequence files were rotated, or inverted and rotated, and renumbered so that their first bp corresponds to that of MG1655.

Eleven genomes have significant inversions, translocations or both relative to the consensus organization. To reduce the complexity of the multiple-alignment, the following inversions and translocations were corrected manually to the same orientation and/or position as in MG1655 using Clone Manager (Scientific and Educational Software): 4.2-5.3 kbp inversions between comparable pairs of oppositely oriented asparagine tRNA sequences in the five B2 strains IHE3034, ABU 83792, CFT073, 536 and SE15; a 1.2 Mbp inversion between oppositely oriented IS3 insertions in P12b; a 92-kbp inversion between oppositely oriented IS629 insertions in O111:H-; a 1.4 Mbp inversion between oppositely oriented IS110 insertions in SMS-3-5; an 8.8 kbp inversion of uncertain origin in ED1A; a 456 kbp inversion with five translocations at a single IS621 insertion in O103:H2; and an inversion of 471 kbp between oppositely oriented IS621 insertions plus a 5.3-kbp inversion at a single IS621 insertion in O26:H11. The O26:H11 genome in the multiple alignment retained two substantial translocations relative to MG1655.

In analyzing scattered SNP clusters in the six least-diverged genome-pairs, we found that the sequence files of CFT073 and APEC O1 have clusters of sequence-ambiguity codes that were being scored as SNPs. Such SNPs were eliminated from all genomes in our analyses simply by requiring that each SNP contain only A, G, C or T.

Basic genomes of individual strains are 94.5% to 99.4% of the total length of the filtered multiple-alignment (due to deletions) and are 68.7% to 84.4% of their corresponding complete-genome lengths (Table S3), reflecting different percentages occupied by mobile elements and exchangeable operons or functional modules present in fewer than 22 of these genomes. Some exchangeable operons or gene clusters that should be considered part of the basic genome did not have enough representation in this set of genomes to have passed the filter.

## SI Computational model of divergence

In our numerical simulations we follow the divergent evolutionary path of two bacterial strains within a population of co-evolving strains with a constant effective population size $N_e$. Computational limitations do not allow us to explicitly model several billions of other strains in the population. Instead we use the mathematical expressions derived below to describe the diversity within the population. We subsequently simulate a Markov chain model following the time course of sequence divergence in a pair of strains. In agreement with the empirical data on *E. coli's* basic genome, each of two strains in our simulation has a circular genome comprising $N_G = 4000$ gene-sized segments consisting of $L_G = 1000$ base pairs each. To speed up our simulations, the finest resolution of genomes in our model is at the level of a segment rather than at the level of individual base pairs.

Our simulations aim at modeling the neutral evolution of the basic genome by random point mutations and homologous recombinations. Thus, they ignore evolutionary processes rearranging gene order or changing the size of the genome such as inversions, short indels or large-scale additions and deletions of genomic fragments. In this simplified picture, in each generation one of two things may happen: a) a randomly selected segment acquires a point mutation a rate $\mu$ per base per generation or b) a randomly selected segment on the genome is chosen as the starting point of a horizontal transfer event at a rate $\rho$ per base pair per generation. Horizontal transfer comprises transfer of a genomic region of length $L$ segments from a randomly selected donor strain from the population of co-evolving strains. Similar to the model used in (3), we assume that segments are indivisible in horizontal transfer. The length of the transferred region is chosen from an exponential distribution with mean $L_t = \langle L \rangle$.

We next describe how we chose the donor strain of the horizontally transferred segment. It is known that the probability of successful transport followed by homologous recombination of a donor fragment into the recipient genome, which we refer to as transfer efficiency, decreases with the increasing sequence divergence between the donor and recipient genome regions. See for example (16, 35-37). Factors likely to affect transfer efficiency include: 1) ecological or geographical niche separation of donor and recipient strains; 2) likelihood that generalized transducing phages from the donor population are capable of infecting recipient strains; 3) defense mechanisms such as DNA Restriction Modification (RM) systems; 4) efficiency of integration into the recipient genome by homologous recombination; and 5) probability that an integrated fragment becomes fixed in the population. In our simulations we assume that transfer efficiency can be approximated by an exponentially decreasing probability of successful recombinant transfer, $p \propto e^{-\frac{\delta}{\delta_{TE}}}$, where $\delta$ is the local sequence divergence between the transferred fragment (comprising on average $L_t$ segments) in donor and recipient genomes. $\delta_{TE}$ is the parameter that dictates the severity of the penalty. For example, a higher $\delta_{TE}$ implies lowered restriction to transfer and vice versa. Previous modeling efforts have also used an exponential biasing function to favor horizontal transfer among closely related strains (3, 38).

**Theoretical framework**

In our computational analysis, we are interested in understanding how a pair of strains, referred to as strains X and Y from here onward, acquires SNPs as the two genomes diverge from their common ancestor as members of the recombining population of $N_e$ strains. We have developed a semi-analytical theoretical model to estimate the divergence among different segments of strains X and Y by taking into account transfers from and within the rest of the population.

Let us denote by $\bar{\delta}(t) = \{\delta_1(t), \delta_2(t), \dots, \delta_{4000}(t)\}$ the vector representing the local sequence divergence in all 4000 segments as a function of time $t$ of divergence from the last common ancestor of X and Y. Local sequence divergence in a given segment is simply the number of SNPs in that segment divided by 1000, the length of each segment. In what follows we analytically derive the transition probability matrix of a Markov process to propagate $\bar{\delta}(t)$ probabilistically in time.

In every generation, the vector of divergence $\bar{\delta}(t)$ is updated if a) there is a point mutation in one of the segments in one of the two genomes (happens at a rate $\mu$ per segment per generation) or b) there is a horizontal transfer event starting at one of the segments in one of the genomes (happens at $\rho$ per segment per generation (diagrammed below). Updating the SNPs vector after a point mutation is straightforward: a mutation in $k^{th}$ segment will increase the $k^{th}$ component of $\bar{\delta}(t)$, $\delta_k(t)$, by 1/1000. On the other hand, horizontal transfer of a genomic +fragment comprising $L_t$ segments in one of the two compared strains will most likely acquire foreign genomic fragments from a randomly chosen strain (referred to as strain D, for donor, from here onward) from the population of co-evolving strains. This horizontal transfer will update the divergence vector at $l_t$ consecutive segments at the same time. Without loss of generality, if strain X receives a horizontal transfer from a donor strain D, the local divergence between X and Y in the horizontally transferred region after the transfer will be the local divergence in the same region between the donor D and Y.

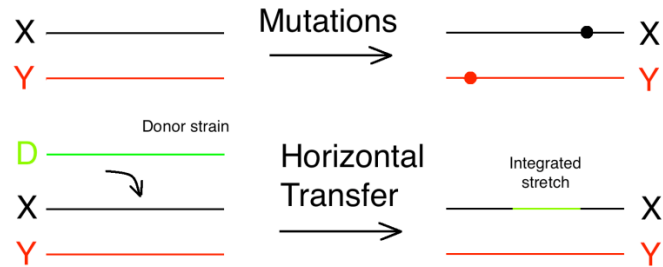

We model horizontal transfer as a two-step process. First, a donor genomic fragment from strain D in the co-evolving population is chosen at random and second, the integration of this fragment into the genome of strain X is attempted with the probability of success determined by transfer efficiency: $p \propto e^{-\frac{\delta}{\delta_{TE}}}$. Given that horizontally transferred fragments have variable lengths, overlap between different horizontal transfer events is possible. In this instance, the phylogenetic lineage of a given fragment cannot be uniquely established since different segments comprising the fragment may potentially have come from different strains. Nevertheless, as a first approximation, justified *a posteriori* by the agreement of the model with empirical observations, we assume that the coalescent framework (see below), which strictly speaking is applicable only to indivisible fragments, is still a valid approximation in tracing the lineage of transferred DNA. Consequently, we derive the conditional probability, $p(\delta_a|\delta_b)$ for a typical fragment that is a candidate for horizontal transfer. Here $\delta_b$ (b for before) denotes the local divergence between X and Y before the horizontal transfer and $\delta_a$ (a for after) denotes the local divergence after integration of the transferred region in X.

A genome fragment integrated in X by recombinant transfer from D will have three possible phylogenetic relationships with the equivalent fragment in the diverging lineage Y, relative to their most recent common ancestor (MRCA): 1) D is more closely related to X than to Y; 2) D is more closely related to Y than to X; or 3) X and Y are more closely related to each other than either is to D.

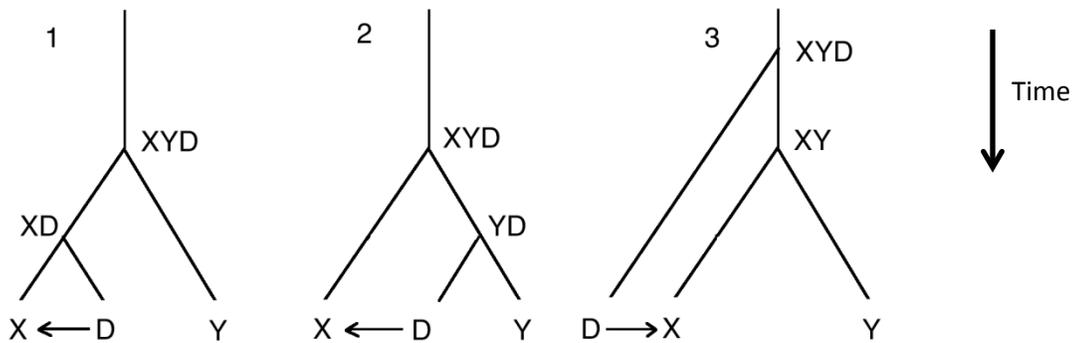

**Case 1:** Here, we want to find the probability $p(\delta_a = \delta_b | \delta_b)$ when the common ancestor XD of X and D is younger than XYD, the common ancestor of X, Y, and D. Note that in this case, $\delta_{XD} < \delta_{XY} = \delta_b$ and $\delta_a = \delta_b$.

Assume that the number of generations separating X and Y is $s_{XY}$. Let D be a randomly chosen strain. The probability that D and X have their MRCA $s_{XD}$ generations ago can be found as follows (39-40). We start with D and choose an ancestor from the previous generation of strains randomly and keep going back in time till we coalesce with the ancestral lineage of X. At every generation, X has a unique ancestor (among $N_e$ strains in the population at that time). At any generation, the probability that we won't randomly pick an ancestor for D that is also in the ancestral lineage of X is $1 - \frac{1}{N_e}$. Consequently the probability that D will coalesce with the ancestral lineage of X in exactly $s_{XD}$ generations is = $\frac{1}{N_e}\left(1 - \frac{1}{N_e}\right)^{s_{XD}-1} \approx \frac{1}{N_e} e^{-\frac{s_{XD}}{N_e}}$.

Similarly, the probability that D avoids the ancestral lineage of Y before coalescing with XD is equal to $= \left(1 - \frac{1}{N_e}\right)^{s_{XD}-1} \approx e^{-\frac{s_{XD}}{N_e}}$. Combining the two, the probability that X and D are separated by $s_{XD}$ generations and XYD is older than XD is given by the product of the two, $p = \frac{1}{N_e} e^{-\frac{2s_{XD}}{N_e}}$.

We can convert the above probability into divergences by noting $\delta = 2\mu s$ and by defining $\theta = 2\mu N_e$ as the average local divergence in the population. We have $p = \frac{1}{\theta} e^{-\frac{2\delta_{XD}}{\theta}}$. Finally integrating over all possible $\delta_{XD} < \delta_{XY} = \delta_b$ by applying the transfer efficiency criterion,

$$p(\delta_a = \delta_b | \delta_b) = \frac{1}{\Omega} \int_0^{\delta_{XY}=\delta_b} e^{-\frac{\delta_{XD}}{\delta_{TE}}} \times \frac{1}{\theta} e^{-\frac{2\delta_{XD}}{\theta}} d\delta_{XD} \qquad (1)$$

Here $\Omega$ is an overall normalization constant.

**Case 2:** Here, we want to find out the probability $p(\delta_a, \delta_a < \delta_b | \delta_b)$ when YD, MRCA of Y and D, is younger than XYD. Note in this case that $\delta_{XD} = \delta_b$ and $\delta_{DY} = \delta_a < \delta_b$.

Similar to case 1, the probability that Y and D had their MRCA exactly $s_{DY}$ generations ago is given by $p = \frac{1}{N_e} e^{-\frac{s_{DY}}{N_e}}$. The probability that D does not coalesce with the ancestral lineage of X before coalescing with that of Y is given by $p = e^{-\frac{s_{DY}}{N_e}}$ (see case 1). The product of the two gives the probability, after including transfer efficiency criterion, (converting to divergences)

$$p(\delta_a, \delta_a < \delta_b | \delta_b) = \frac{1}{\Omega} \frac{1}{\theta} e^{-\frac{2\delta_a}{\theta}} \times e^{-\frac{\delta_b}{\delta_{TE}}} \qquad (2)$$

**Case 3:** Here we want to find out the probability $p(\delta_a, \delta_a > \delta_b|\delta_b)$ when the MRCA of X and D and the MRCA of Y and D is the same (XYD) and is older than the MRCA of X and Y, XY. Note in this case that $\delta_{DY} = \delta_{XD} = \delta_a > \delta_b = \delta_{XY}$.

In this case, we first require that D does not coalesce with the ancestral lineage of X and of Y in $s_{XY}$ generations. This probability is given by $p = e^{-\frac{2s_{XY}}{N_e}}$ (see case 1 and 2). After avoiding the ancestral lineages for the first $s_{XY}$ generations, we now require D to coalesce with the ancestral lineages of both X and Y in exactly $s_{DY} - s_{XY}$ generations. This probability is given by $p = \frac{1}{N_e}e^{-\frac{(s_{DY}-s_{XY})}{N_e}}$ (see case 1 and 2). Finally, the product of the two gives the probability that D coalesces with Y in $s_{DY}$ generations and $s_{DY} > s_{XY}$ Combining the two, we get $p = \frac{1}{N}e^{-\frac{(s_{DY}+s_{XY})}{N_e}}$. Converting to divergences, applying the transfer efficiency criterion, we have

$$p(\delta_a, \delta_a > \delta_b|\delta_b) = \frac{1}{\Omega}\frac{1}{\theta}e^{-\frac{\delta_a+\delta_b}{\theta}} \times e^{-\frac{\delta_a}{\delta_{TE}}} \quad (3)$$

The normalization constant $\Omega = \Omega(\delta_b)$ depends on the original local divergence $\delta_b$ between X and Y and is simply defined as the normalization constant that ensures that probabilities in (1), (2), and (3) add up to one.

Combining Eq. (1), (2), and (3), we can finally write down the overall conditional probability $p(\delta_a|\delta_b)$ is given by

$$p(\delta_a|\delta_b) = \frac{1}{\Omega}\Big(Di(\delta_a - \delta_b) \times p(\delta_a = \delta_b|\delta_b) + \Theta(\delta_b - \delta_a) \times p(\delta_a, \delta_a < \delta_b|\delta_b) + \Theta(\delta_a - \delta_b) \times p(\delta_a, \delta_a > \delta_b|\delta_b)\Big) \quad (4)$$

Here, $Di(x)$ is the Dirac Delta function and $\Theta(x)$ is the Heaviside function. Using the conditional probability in Eq. (4) we predict the distribution of the local divergence $\delta_a$ between X and Y *after* horizontal transfer from $\delta_b$, the corresponding local divergence before the horizontal transfer.

Thus, we have analytically derived the probability to update a local fragment of the SNP vector $\bar{\delta}(t)$. Combined with the relative rates of mutation and recombination, we can construct the Markovian transition probability for $\bar{\delta}(t) \to \bar{\delta}(t+1)$. Using this Markovian transition probability, we numerically propagate the stochastic trajectory of $\bar{\delta}(t)$ as detailed below.

**Simulations**

Using Eq. (4), we are equipped to perform a Markov chain simulation to propagate $\bar{\delta}(t)$. We proceed as follows:

1) Initialize the SNPs vector as a vector of zeros at time $t = 0$. This implies that we start with two strains with identical genomes at time $t = 0$
2) At every generation do the following

a. Since we know that $\mu \ll 1$ is very low, it is very unlikely that a single mutation event brings about two or more SNPs. Thus, first draw $G = 4000$ random variables $z_k$ uniformly from [0, 1]. If $z_k < \mu$, we add 1/1000 to $\delta_k(t)$.
b. Draw 4000 random numbers $r$ uniformly from [0, 1] and compare them with $\rho$. Identify segment(s) with $r < \rho$. These segments are candidates for horizontal transfer. Given that the rate of horizontal transfer is very likely to be low, in any time step it is very unlikely that more than one segment would be selected for horizontal transfer.
    i. If segment $k$ is selected for horizontal transfer, draw an integer $L$ from an exponential distribution with the mean $L_t$. Genes from $k$ to $k+L-1$ are potential candidates for horizontal replacement of genomic fragments. Note that the genome is circular i.e. segments 1 and 4000 are neighbors of each other, and that we allow segment transfers to start and end only at segment boundaries.
    ii. Find the local divergence $\delta_b$ in the selected genomic fragment over the $L$ segments before the horizontal transfer and, using the conditional probability $p(\delta_a|\delta_b)$ defined in Eq. (5), sample $\delta_a$.
    iii. Draw $L$ Poisson random variables with parameter $\delta_a$ and distribute them randomly on the $L$ segments and update the SNPs vector for these $L$ segments using the newly sampled Poisson variables.

Step 2.b.iii assumes that the newly acquired genomic fragment has no spatial correlations of its SNPs, an assumption which is reasonable for the current analysis but may need to be relaxed if one is interested in understanding SNP positional correlations along the genome.

Each Markov chain starts at $t=0$ with strains X and Y with identical genomes i.e. $\bar{\delta}(t) = \bar{0}$. As time increases, the global genomic divergence $\Delta$ increases like a random walk with a drift. To generate SNPs vectors $\bar{\delta}(t)$ at any given $\Delta$, we stop the Markov chain when $\Delta$ reaches within 0.01% of the prescribed divergence. For each $\Delta$ we collect 100 such Markov chains. It is a straightforward calculation to generate the figures in the main text from the SNPs vectors.

## Parameters

Our theoretical and numerical model depends on 4 dimensionless parameters:

1) $\theta = 2\mu N_e$, Watterson's measure of effective population diversity
2) $\frac{\rho}{\mu}$, the ratio of the horizontal transfer rate to the mutation rate
3) $\frac{L_t}{L_G \times N_G}$, the fraction of the genome length replaced in a single transfer event
4) $\delta_{TE}$, the transfer efficiency in units of pairwise sequence divergence

A ratio of $\frac{\rho}{\mu} = 0.31$ best fits the profile of average clonal and recombined SNP densities shown in Figures 3b and 3c. Based on a mutation rate $\mu = 8.9 \times 10^{-11}$ per bp per generation (25), that ratio corresponds to a recombination rate of $\rho = 2.8 \times 10^{-11}$ per bp per generation.

The fraction of genome length replaced in a single transfer event dictates the maximum number of uninterrupted strings of transferred segments (Figure 4c). Trial and error showed that $\frac{L_t}{L_G \times N_G} = 0.075\%$ gives a good fit to the data, corresponding to an average transferred length of $L_t = 3$ kbp.

The exponential tail in the SNP distributions generated in our numerical simulations (Fig. 2c) is influenced by the combined parameter $\frac{1}{\theta} + \frac{1}{\delta_{TE}}$ (see Eq. 3). For genomes with a large number of genes, Watterson's $\theta = 2\mu N_e$ is roughly equal to the maximum of the overall divergence $\Delta$ between any two genomes in the population. The maximum divergence of our 32 genomes is 2.6% so θ must at least be greater than 2.6%, which imposes a lower bound on the effective population size $N_e = \theta/2\mu > 1.5 \times 10^8$. If transfers were perfectly efficient ($\delta_{TE} \gg 1$), the slope of the exponential tail would be governed entirely by $\theta$. For $\theta = 2.6\%$, SNP distributions predicted by our model have exponential tails that are about twice wider than those observed in the empirical data (Fig. 2b). Thus, we assumed that the slope of the exponential distribution in the tails of $P(\delta_i)$ is governed primarily by $\delta_{TE}$. To simplify calculations in our model we set $\theta \gg \delta_{TE}$. In this limit the value $\delta_{TE} = 0.8\%$ provides the best fit (Fig. 2c) to the exponential slopes observed in SNP histograms of genome-pairs within group A and between group A and B1 (Fig. 2b).

Comparing our results with previous work, our estimated average transfer length of 3 kbp is considerably longer than the ~50 bp estimated by Touchon et al. (13) or the 542 bp estimated by Didelot et al. (14). The ratio $\frac{\rho}{\mu} = 0.31$ in our model is smaller than the estimate of $0.0128/0.0125 \cong 1$ by Didelot et al. (14) (albeit with a much shorter length of recombined regions) and much smaller than the 2.47 calculated by Touchon et al. (13). A composite parameter $\frac{r}{m} = \frac{\rho}{\mu} L_T \delta_{TE}$, an estimate of the relative contributions of recombinant transfers and random mutations to overall divergence, is 11.2 in our model, much lower than the early estimate of 50 by Guttman and Dykhuizen (20), considerably higher than 0.34 by McNally et al. (21) and 1.5 by Touchon et al. (13), and somewhat higher than 7, estimated by Didelot et al. (14).